\documentclass[aps,twocolumn,longbibliography,nofootinbib]{revtex4-2}
\usepackage[utf8]{inputenc}
\usepackage[T1]{fontenc}
\usepackage{amsmath}
\usepackage{amssymb}
\usepackage{amsthm}
\usepackage{siunitx}
\usepackage{physics}
\usepackage{graphicx}
\usepackage{subfigure}
\usepackage{tabularx}
\usepackage{booktabs}
\usepackage{multirow}
\usepackage{hyperref}

\hypersetup{hypertex=true, colorlinks=true, linkcolor=blue, urlcolor=red, citecolor=red}

\begin{document}

\title{Building ground states of Hubbard model by time-ordered bound-pair
injection}
\author{K. L. Zhang}
\author{Z. Song}
\email{songtc@nankai.edu.cn}
\affiliation{School of Physics, Nankai University, Tianjin 300071, China}

\begin{abstract}
According to energy band theory, ground states of a normal conductor and
insulator can be obtained by filling electrons individually into energy
levels, without any restrictions. It fails when the electron-electron
correlation is taken into account. In this work, we investigate the dynamic
process of building ground states of a Hubbard model. It is based on
time-ordered quantum quenches for unidirectional hopping across a central
and an auxiliary Hubbard model. We find that there exists a set of optimal
parameters (chemical potentials and pair binding energy) for the auxiliary
system, which takes the role of electron-pair reservoir. The exceptional
point dynamics in non-Hermitian quantum mechanics allows the perfect
transfer of electron pair from the reservoir to the central system,
obtaining its ground states at different fillings. The dynamics of
time-ordered pair-filling not only provides a method for correlated quantum
state engineering, but also reveals the feature of the ground state in an
alternative way.
\end{abstract}

\maketitle

\section{Introduction}

Understanding the quantum states of strongly interacting many-body systems
via quantum dynamics is one of the promising methods in contemporary
condensed matter physics. Compared to weakly or noninteracting systems,
strong interactions can induce fascinating phenomena, which cannot be
understood by conventional band theory. One example is the Mott insulating
state: for the ground state of a fermionic lattice system with a half-filled
band, strong interactions can make this system insulating \cite%
{hubbard1963electron, imada1998metal, staudt2000phase} from the conducting
state. The essence of what happens is correlation between two fermions with
opposite spins. High-temperature superconductivity, as another example, can
arise from the correlated motion of holes in an antiferromagnetic Mott
insulator \cite{anderson1987resonating, lee2006doping}. Theoretically, such
a correlation can be characterized by the correlation function. However, it
is a challenge to measure the correlation function in the experiment \cite%
{parsons2016site}. Recently, correlated insulator has attracted much
attention due to the discovery of twisted bilayer graphene \cite%
{bistritzer2011moire, cao2016superlattice, kim2017tunable,
cao2018correlated, cao2018unconventional, yankowitz2019tuning,
lu2019superconductors}.

A conceptually clear and frequently used dynamic approach is a quantum
quench, where one starts in the ground state of a given Hamiltonian and then
suddenly changes the parameters of this Hamiltonian. After a sufficiently long
time, the evolved state may deviate from the ground state and is not the
eigenstate of the quenched Hamiltonian in general, since the process is
nonadiabatic. Nevertheless, it is expected that the nonequilibrium state
contains both the information of initial and final Hamiltonians. Many
research efforts have been devoted to this subject \cite{calabrese2006time,
manmana2007strongly, iucci2009quantum, schiro2011quantum,
moeckel2008interaction, eckstein2009thermalization,kollath2007quench,
biroli2010effect, greiner2002collapse, kinoshita2006quantum,
trotzky2012probing, cheneau2012light, gring2012relaxation}.

In normal conductors and insulators, the situation is well described by free
electron theory, where the electrons behave as free particles. Viewed in
this context, electrons can be injected into or emitted from the material
individually. Technically,\ the ground state of free electron gas\ can be
built by a\ dynamic process in the framework of quantum mechanics. A natural
question is how, in the presence of interaction between electrons with
opposite spins, a correlated many-body ground state is formed dynamically.
Despite being a great computational challenge for simulating the dynamics in
large size quantum many-body systems, theoretical calculations on small
sized systems may provide new insights into the experimental observations.
Recent advances in quantum simulations of the Hubbard model with ultracold
atoms have offer a multifunctional platform to unveil properties of the
strongly correlated system \cite{syassen2008strong, bakr2009quantum,
zhu2014suppressing, parsons2015site, cheuk2015quantum, cheuk2016observation,
parsons2016site, tomita2017observation, sponselee2018dynamics,
sato2019light, esslinger2010fermi, booker2020non, zhang2020dynamic}.

In this work, we propose a method to build ground states of a Hubbard model
in the framework of quantum dynamics. It is an extension of the conventional
quenching method, switching a system from $N$-site to $(N+1)$-site lattices.
We focus on an\ $N$-site\ Hubbard model with a side-coupled site. Initially,
the central system is set in ground state at a certain filling and the
side-coupled site is empty. We consider two types of side-coupling strength---Hermitian and non-Hermitian \cite{bender1998real, mostafazadeh2002pseudo,
jin2010physics, muller2008exceptional, heiss2012physics, rotter2015review}
ones. In the Hermitian case, analytical analysis and numerical simulation show
that there exist optimal parameters, chemical potential\ and pair binding
energy, for the extra site, with which the time evolution exhibits perfect
oscillation, indicating that a bound pair of electrons can be extracted and
returned back to the central system coherently. In the non-Hermitian case
with unidirectional hopping, the exceptional point (EP) \cite%
{muller2008exceptional, heiss2012physics, rotter2015review} dynamics allows
the complete transfer of electron pair between the central system and the
extra site. Furthermore, we investigate the dynamic process of building ground
states of a Hubbard model. It bases on time-ordered quantum quenches for
unidirectional hopping across a central and an auxiliary Hubbard model. Finally, we provide a scanning scheme to determine the quenching
parameters. Numerical simulation for small-size one-dimensional
($1$D) and two-dimensional ($2$D) systems shows that the ground states at
different fillings can be built coherently with high fidelity. Our finding
not only provides a method for correlated quantum state engineering, but
also reveals the feature of the ground state in an alternative way.

This paper is organized as follows. In Sec. \ref{Model and single-pair
oscillation}, we introduce the model Hamiltonian and study the dynamics of
single-pair oscillation. In Sec. \ref{Complete pair transport}, we
investigate the dynamics of complete pair transport by a non-Hermitian method.
In Sec. \ref{Time-ordered bound-pair injection}, we propose a dynamical way
to build the ground state of the Hubbard model by time-ordered bound-pair
injection. In Sec. \ref{Determinations}, we provide a scanning
scheme to determine the quenching parameters. Finally, we summarize our
results in Sec. \ref{Summary}.

\section{Model and single-pair oscillation}

\label{Model and single-pair oscillation}

We consider the Hubbard model $H_{\mathrm{c}}$ connected with a side-coupled
site $\mathrm{p}$. The Hamiltonian%
\begin{equation}
H=H_{\mathrm{c}}+H_{\mathrm{p}}
\end{equation}%
consists of two parts. Here $H_{\mathrm{c}}$ is a simple Hubbard model on an 
$N$-site bipartite lattice with equal sites of two sublattices,
\begin{equation}
H_{\mathrm{c}}=\sum_{l>l^{\prime }}\sum_{\sigma=\uparrow ,\downarrow}\left(
J_{l,l^{\prime }}c_{l,\sigma }^{\dag }c_{l^{\prime },\sigma }+\mathrm{H.c.}%
\right) +\sum_{l}U_{l}n_{l,\uparrow }n_{l,\downarrow },  \label{Hc}
\end{equation}%
where the operator $c_{l,\sigma }$ ($c_{l,\sigma }^{\dagger }$) is the
usual annihilation (creation) operator of an electron with spin $\sigma \in
\left\{ \uparrow ,\downarrow \right\} $ at site $l$, and $n_{l,\sigma
}=c_{l,\sigma }^{\dagger }c_{l,\sigma }$ is the number operator for a
particle of spin $\sigma $ on site $l$. The hopping $%
J_{l,l^{\prime }}$ and interaction $U_{l}$ are required to be real; the
system can be divided into two sublattices $\mathrm{A}$ and $\mathrm{B}$
such that $J_{l,l^{\prime }}=0$ whenever $l,l^{\prime }\in \{\mathrm{A}\}$
or $l,l^{\prime }\in \{\mathrm{B}\}$, and the dimension of the lattice is
not yet assumed. The side-coupled term is 
\begin{eqnarray}
H_{\mathrm{p}} &=&J_{\mathrm{p}}\sum_{\sigma =\uparrow ,\downarrow
}c_{\alpha ,\sigma }^{\dagger }c_{\mathrm{p},\sigma }+\text{\textrm{H.c.}}%
+U_{\mathrm{b}}n_{\mathrm{p},\uparrow }n_{\mathrm{p}\mathbf{,}\downarrow } 
\notag \\
&&+\mu _{\uparrow }n_{\mathrm{p},\uparrow }+\mu _{\downarrow }n_{\mathrm{p}%
\mathbf{,}\downarrow },
\end{eqnarray}%
where $c_{\mathrm{p},\sigma }$\ is the annihilation operator of an electron
with spin $\sigma $\ on the side-coupled site $\mathrm{p}$; $\alpha \in \{%
\mathrm{A},\mathrm{B}\}$; $\mu _{\sigma }$\ is the chemical potential of the
electron with spin $\sigma $ and $U_{\mathrm{b}}$\ is the binding energy of
electron pair on the $\mathrm{p}$ site. In this work, we mainly consider the
case with $N$-electron filling.

We first review some well-known model properties of the Hubbard model $H_{%
\mathrm{c}}$\ that are crucial to our conclusion. At first, $H_{\mathrm{c}}$%
\ possesses SU(2) symmetry 
\begin{equation}
\left[ s^{\pm },H_{\mathrm{c}}\right] =\left[ s^{z},H_{\mathrm{c}}\right] =0,
\end{equation}%
with $s^{+}=\left( s^{-}\right) ^{\dagger }=\sum_{l}s_{l}^{+}$ and $%
s^{z}=\sum_{l}s_{l}^{z}$, where the local operators $s_{l}^{+}=c_{l,\uparrow
}^{\dagger }c_{l,\downarrow }$ and $s_{l}^{z}=\left( n_{l,\uparrow
}-n_{l,\downarrow }\right) /2$ obey the Lie algebra, that is 
\begin{equation}
\lbrack s_{l}^{+},s_{l}^{-}]=2s_{l}^{z},[s_{l}^{z},s_{l}^{\pm }]=\pm
s_{l}^{\pm }.
\end{equation}%
Secondly, $H_{\mathrm{c}}$ has spin reversal symmetry defined by%
\begin{equation}
\mathcal{T}H_{\mathrm{c}}\mathcal{T}^{-1}=H_{\mathrm{c}},
\end{equation}%
where $\mathcal{T}$ is the spin reversal operator with the action $\mathcal{T}%
c_{l,\uparrow }\mathcal{T}^{-1}=c_{l,\downarrow }$\ and $\mathcal{T}%
c_{l,\downarrow }\mathcal{T}^{-1}=c_{l,\uparrow }$ for all $l$. Then the
eigenstates of $H_{\mathrm{c}}$ are also the eigenstates of operators $s$
and $s^{z}$. And any eigenstates with nonzero $s$\ must be degenerate. In
addition, according to Lieb's theorem \cite{lieb1989two} for a bipartite
lattice, in the repulsive case $U_{l}=U>0$, the ground state of $H_{\mathrm{c%
}}$\ at half filling is unique and has spin $s=0$. The following
considerations are based on these properties.

We are interested in the lowest energy eigenstates of $H_{\mathrm{c}}$\ in
the invariant subspaces with the numbers of electrons $n$ ($s^{z}=0$), $n-1$
($s^{z}=\pm 1/2$), and $n-2$ ($s^{z}=0$), respectively. Here $n$ is set to
an even number satisfying $2\leq n\leq N$. The corresponding Schr\"{o}dinger
equations are%
\begin{eqnarray}
&&\left[ H_{\mathrm{c}}-E_{\mathrm{g}}(n,0)\right] \left\vert \psi _{\mathrm{%
g}}(n,0)\right\rangle =0,  \notag \\
&&\left[ H_{\mathrm{c}}-E_{\mathrm{g}}(n-1,\pm 1/2)\right] \left\vert \psi _{%
\mathrm{g}}(n-1,\pm 1/2)\right\rangle =0,  \notag \\
&&\left[ H_{\mathrm{c}}-E_{\mathrm{g}}(n-2,0)\right] \left\vert \psi _{%
\mathrm{g}}(n-2,0)\right\rangle =0,
\end{eqnarray}%
where $E_{\mathrm{g}}$\ is the ground-state energy in each invariant
subspace, and $E_{\mathrm{g}}(n-1,1/2)$ $=E_{\mathrm{g}}(n-1,-1/2)$
according to symmetries in the above analysis.

Now we consider the system with $H$ in the subspace spanned by the basis set%
\begin{eqnarray}
\left\vert 1\right\rangle &=&\left\vert \psi _{\mathrm{g}}(n,0)\right\rangle
\left\vert 0\right\rangle _{\mathrm{p}},  \notag \\
\left\vert 2\right\rangle &=&\left\vert \psi _{\mathrm{g}}(n-1,-1/2)\right%
\rangle \left\vert \uparrow \right\rangle _{\mathrm{p}},  \notag \\
\left\vert 3\right\rangle &=&\left\vert \psi _{\mathrm{g}}(n-1,1/2)\right%
\rangle \left\vert \downarrow \right\rangle _{\mathrm{p}},  \notag \\
\left\vert 4\right\rangle &=&\left\vert \psi _{\mathrm{g}}(n-2,0)\right%
\rangle \left\vert \uparrow \downarrow \right\rangle _{\mathrm{p}},
\label{basis}
\end{eqnarray}%
where $\left\vert 0\right\rangle _{\mathrm{p}}$\ is the vacuum state of the 
side-coupled site and $\left\vert \sigma \right\rangle _{\mathrm{p}}=c_{%
\mathrm{p},\sigma }^{\dag }\left\vert 0\right\rangle _{\mathrm{p}}$,\ $%
\left\vert \uparrow \downarrow \right\rangle _{\mathrm{p}}=c_{\mathrm{p}%
,\uparrow }^{\dag }c_{\mathrm{p},\downarrow }^{\dag }\left\vert
0\right\rangle _{\mathrm{p}}$. Based on the basis set, the matrix
representation of the effective Hamiltonian for system $H$ is%
\begin{equation}
h=\left( 
\begin{array}{cccc}
\epsilon _{1} & \kappa _{1} & \kappa _{1} & 0 \\ 
\kappa _{1} & \epsilon _{2} & 0 & \kappa _{2} \\ 
\kappa _{1} & 0 & \epsilon _{3} & \kappa _{2} \\ 
0 & \kappa _{2} & \kappa _{2} & \epsilon _{4}%
\end{array}%
\right) ,  \label{h_hermitian}
\end{equation}%
where the matrix elements 
\begin{eqnarray}
\epsilon _{1} &=&E_{\mathrm{g}}(n,0),\epsilon _{2}=E_{\mathrm{g}%
}(n-1,1/2)+\mu _{\uparrow },  \notag \\
\epsilon _{3} &=&E_{\mathrm{g}}(n-1,1/2)+\mu _{\downarrow },  \notag \\
\epsilon _{4} &=&E_{\mathrm{g}}(n-2,0)+\mu _{\uparrow }+\mu _{\downarrow
}+U_{\mathrm{b}},  \notag \\
\kappa _{1} &=&\langle 1|H_{\mathrm{p}}|2\rangle =\langle 1|H_{\mathrm{p}%
}|3\rangle =\langle 2|H_{\mathrm{p}}|1\rangle =\langle 3|H_{\mathrm{p}%
}|1\rangle ,  \notag \\
\kappa _{2} &=&\langle 2|H_{\mathrm{p}}|4\rangle =\langle 3|H_{\mathrm{p}%
}|4\rangle =\langle 4|H_{\mathrm{p}}|2\rangle =\langle 4|H_{\mathrm{p}%
}|3\rangle .
\end{eqnarray}%
In this framework, parameters $\left\{ \mu _{\uparrow },\mu _{\downarrow
},U_{\mathrm{b}}\right\} $ determine the dynamics of electron transport. We
consider the following three situations.

(i) $\left\vert \mu _{\downarrow }\right\vert \gg \mu _{\uparrow }=E_{%
\mathrm{g}}(n,0)-E_{\mathrm{g}}(n-1,1/2)$; matrix $h$\ reduces to%
\begin{equation}
h_{\uparrow }=\kappa _{1}\left( \left\vert 1\right\rangle \left\langle
2\right\vert +\left\vert 2\right\rangle \left\langle 1\right\vert \right)
+\varepsilon _{1}\left( \left\vert 1\right\rangle \left\langle 1\right\vert
+\left\vert 2\right\rangle \left\langle 2\right\vert \right) .
\end{equation}

(ii) $\left\vert \mu _{\uparrow }\right\vert \gg \mu _{\downarrow }=E_{%
\mathrm{g}}(n,0)-E_{\mathrm{g}}(n-1,1/2)$; matrix $h$\ reduces to%
\begin{equation}
h_{\downarrow }=\kappa _{1}\left( \left\vert 1\right\rangle \left\langle
3\right\vert +\left\vert 3\right\rangle \left\langle 1\right\vert \right)
+\varepsilon _{1}\left( \left\vert 1\right\rangle \left\langle 1\right\vert
+\left\vert 3\right\rangle \left\langle 3\right\vert \right) .
\end{equation}%
In these two cases, one of the polarized electrons is excluded at the $\mathrm{p}$
site. The dynamics is single-electron oscillation, realizing perfect
transport from the center system to the $\mathrm{p}$ site. We are interested in
the third case, where spin-up and spin-down electrons are all in resonance.

\begin{figure}[tbh]
\centering
\includegraphics[width=0.5\textwidth]{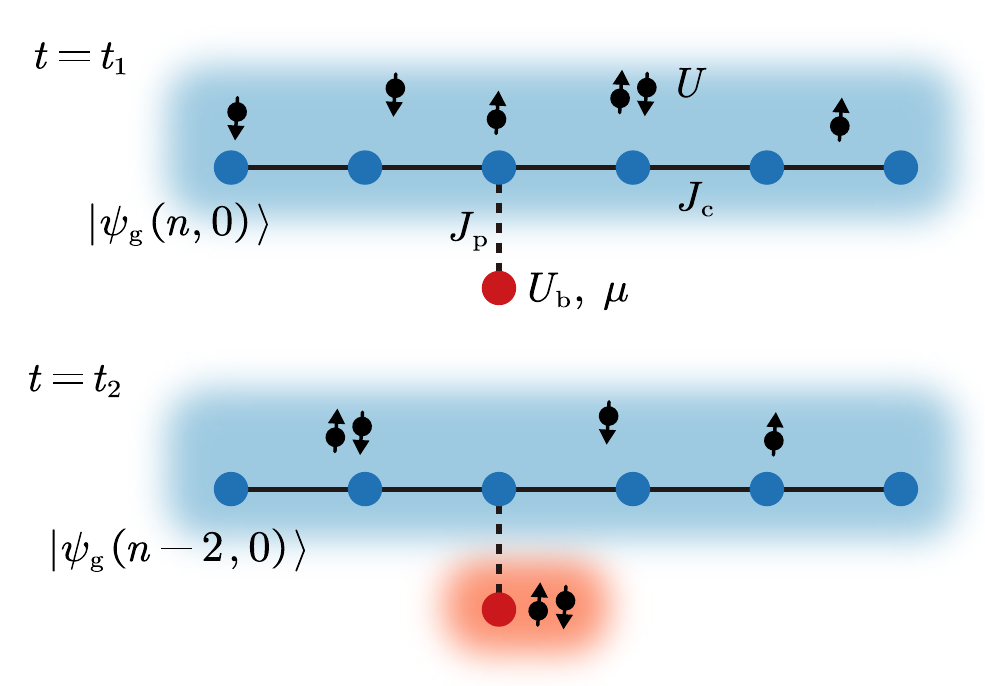}
\caption{Schematic of the complete pair resonance in the Hubbard chain. At
instant $t=t_{1}$, the center system (blue region) is prepared in $n$%
-particle ground state and the side-coupled site (red dot) is empty. With
certain pair binding energy $U_{\text{b}}$ and chemical potential $\protect%
\mu$ [Eq. (\protect\ref{Ub})] of the side-coupled site, an electron pair is
resonantly transmitted to the side-coupled site, and the center system
remains in the $(n-2)$-particle ground state at instant $t=t_{2}$.}
\label{fig1}
\end{figure}

(iii) When taking the parameters as 
\begin{eqnarray}
\mu _{\uparrow } &=&\mu _{\downarrow }=\mu =E_{\mathrm{g}}(n,0)-E_{\mathrm{g}%
}(n-1,1/2),  \notag \\
U_{\mathrm{b}} &=&2E_{\mathrm{g}}(n-1,1/2)-E_{\mathrm{g}}(n-2,0)-E_{\mathrm{g%
}}(n,0),  \label{Ub}
\end{eqnarray}%
matrix $h$\ reduces to%
\begin{equation}
h_{\mathrm{R}}=\left( 
\begin{array}{cccc}
0 & \kappa _{1} & \kappa _{1} & 0 \\ 
\kappa _{1} & 0 & 0 & \kappa _{2} \\ 
\kappa _{1} & 0 & 0 & \kappa _{2} \\ 
0 & \kappa _{2} & \kappa _{2} & 0%
\end{array}%
\right) +E_{\mathrm{g}}(n,0).
\end{equation}%
It supports periodic dynamics with period $T=2\pi /\varepsilon $ ($%
\varepsilon =\sqrt{2\left( \kappa _{1}^{2}+\kappa _{2}^{2}\right) }$), since
the energy levels are always equally spaced $\varepsilon $, and importantly
allows the time evolution%
\begin{eqnarray}
&&\left\vert \psi (t)\right\rangle =e^{-ih_{\mathrm{R}}t}\left\vert
1\right\rangle =\frac{1}{2\left( \kappa _{1}^{2}+\kappa _{2}^{2}\right) }%
\times  \notag \\
&&\{\left( 2\kappa _{1}^{2}\cos \left( \varepsilon t\right) +2\kappa
_{2}^{2}\right) \left\vert 1\right\rangle -i\varepsilon \kappa _{1}\sin
\left( \varepsilon t\right) \left( \left\vert 2\right\rangle +\left\vert
3\right\rangle \right)  \notag \\
&&+\left( 2\kappa _{2}\kappa _{1}\cos \left( \varepsilon t\right) -2\kappa
_{2}\kappa _{1}\right) \left\vert 4\right\rangle \}
\end{eqnarray}%
for the initial state $\left\vert \psi _{\mathrm{g}}(N,0)\right\rangle
\left\vert 0\right\rangle _{\mathrm{p}}$. It demonstrates a pair oscillation
between the center system and the side-coupled site, indicating that a pair
of electrons can be extracted from the ground state of $H_{\mathrm{c}}$\ at
instant $\left( m+1/2\right) T$ ($m=0,1,2,3,...$)\ in the small $J_{\mathrm{p%
}}$\ limit. The maximal pair transport is%
\begin{equation}
\text{Max}\left[ \left\vert \left\langle 4\right. \left\vert \psi
(t)\right\rangle \right\vert ^{2}\right] =\frac{4\left( \kappa _{2}\kappa
_{1}\right) ^{2}}{\left( \kappa _{1}^{2}+\kappa _{2}^{2}\right) ^{2}},
\end{equation}%
which turns to unit at $\kappa _{1}=\kappa _{2}$, indicating complete pair
transport. The schematic illustration of this process for a $1$D system is
shown in Fig. \ref{fig1}, although the above analysis is not
limited to 1D.

\begin{figure*}[tbh]
\centering
\includegraphics[width=1\textwidth]{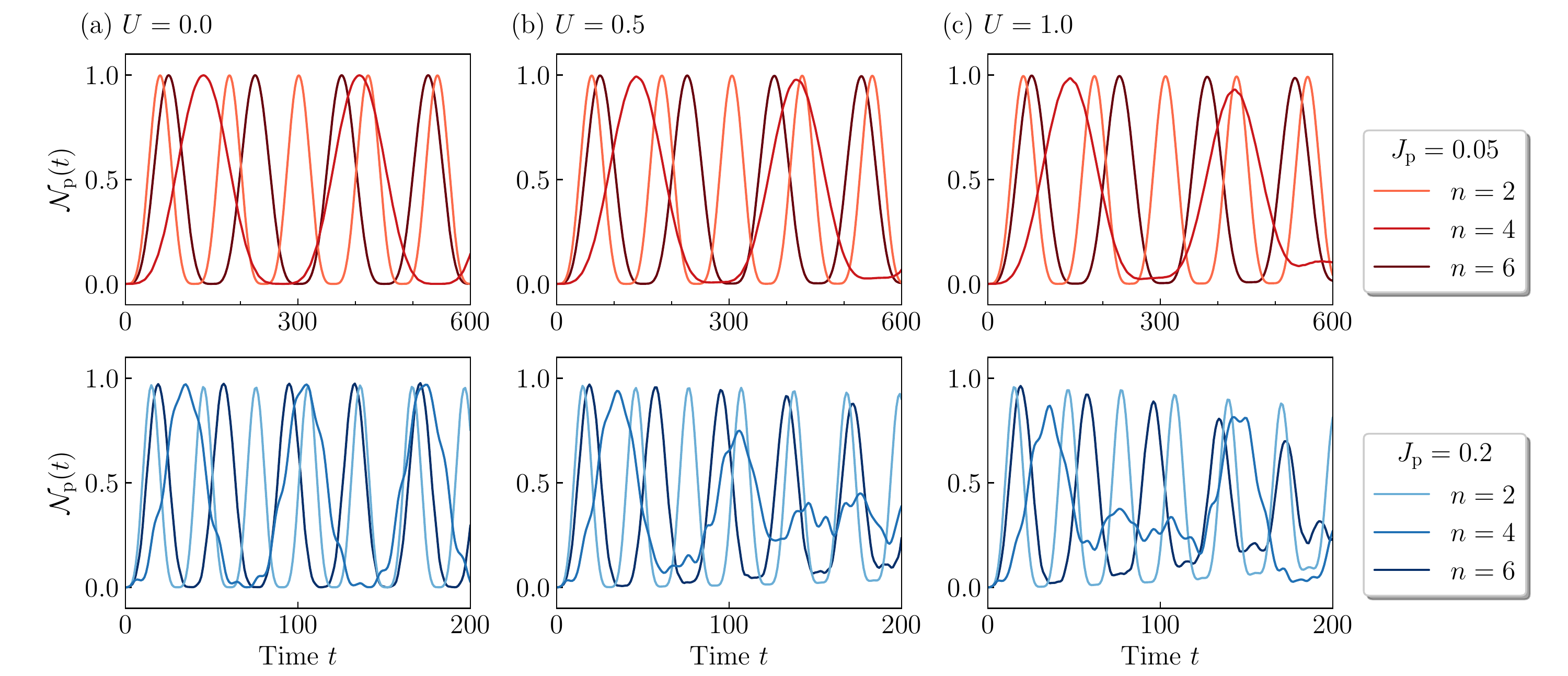}
\caption{Numerical results of the electron pair oscillation for different
fillings $n$ and values of $U_{l}=U$: (a) $U=0$, (b) $U=0.5$, and (c) $U=1.0$%
. The top and bottom panels show the number of electron pairs of the
side-coupled site as a function of time $t$ for coupling strength $J_{%
\mathrm{p}}=0.05$ and $0.2$, respectively. Initially, the center system with
Hubbard interaction $U$ is prepared in the $n$-particle ground state $%
\left\vert \protect\psi _{\mathrm{g}}(n,0)\right\rangle$ of Hamiltonian $H_{%
\mathrm{c}}$ and the side-coupled site is empty. The pair binding energy $U_{%
\text{b}}$ and chemical potential $\protect\mu$ of the side-coupled site are
set to the values defined in Eq. (\protect\ref{Ub}), wherein the
ground-state energies are obtained by the exact diagonalization of the
Hamiltonian $H_{\mathrm{c}}$ in $n$-, $(n-1)$- and $(n-2)$-particle subspaces.
The uniform nearest neighbor hopping is $J_{\mathrm{c}}=1$ and the system
size is $N=6$.}
\label{fig2}
\end{figure*}

For zero $U$, the ground states of $H_{\mathrm{c}}$\ in each invariant
subspace have simple relations 
\begin{eqnarray}
\left\vert \psi _{\mathrm{g}}(n,0)\right\rangle &=&c_{\mathrm{F},\uparrow
}^{\dag }\left\vert \psi _{\mathrm{g}}(n-1,-1/2)\right\rangle  \notag \\
&=&-c_{\mathrm{F},\downarrow }^{\dag }\left\vert \psi _{\mathrm{g}%
}(n-1,1/2)\right\rangle ,
\end{eqnarray}%
and%
\begin{eqnarray}
\left\vert \psi _{\mathrm{g}}(n-1,1/2)\right\rangle &=&c_{\mathrm{F}%
,\uparrow }^{\dag }\left\vert \psi _{\mathrm{g}}(n-2,0)\right\rangle , 
\notag \\
\left\vert \psi _{\mathrm{g}}(n-1,-1/2)\right\rangle &=&c_{\mathrm{F}%
,\downarrow }^{\dag }\left\vert \psi _{\mathrm{g}}(n-2,0)\right\rangle ,
\end{eqnarray}%
where $c_{\mathrm{F},\sigma }^{\dag }$\ is the creation operator of an electron
at Fermi level $\varepsilon _{\mathrm{F}}$. Accordingly, we have 
\begin{eqnarray}
E_{\mathrm{g}}(n,0) &=&\varepsilon _{\mathrm{F}}+E_{\mathrm{g}}(n-1,\pm 1/2)
\notag \\
&=&2\varepsilon _{\mathrm{F}}+E_{\mathrm{g}}(n-2,0).
\end{eqnarray}%
Obviously, we have $\kappa _{1}=\kappa _{2}$ and $U_{\mathrm{b}}=0$,\ which
results in complete extraction. This can be understood from the following
facts: when $U=0$, all the dynamics of $H$ in each invariant subspace with
fixed electron number is governed by its matrix representation in a 
single-particle subspace. In the single-particle subspace,$\ H_{\mathrm{c}}$%
\ can be written as a diagonal form,%
\begin{equation}
H_{0}=\sum_{j}\varepsilon _{j}(\left\vert j,\uparrow \right\rangle
\left\langle j,\uparrow \right\vert +\left\vert j,\downarrow \right\rangle
\left\langle j,\downarrow \right\vert ),
\end{equation}%
where the single-particle eigenstate with eigenenergy $\varepsilon _{j}$ is
defined as $\left\vert j,\sigma \right\rangle =A_{j,\sigma }^{\dag
}\left\vert \mathrm{Vac}\right\rangle $\ ($\sigma =\uparrow ,\downarrow $)
and $\left\vert \mathrm{Vac}\right\rangle $\ is\ the vacuum state of the 
electron operator, i.e., $c_{l,\sigma }\left\vert \mathrm{Vac}\right\rangle
=0$. For any pair eigenstate 
\begin{equation}
\left\vert \psi (n,0)\right\rangle =\prod_{\left\{ j\right\} }A_{j,\uparrow
}^{\dag }A_{j,\downarrow }^{\dag }\left\vert \mathrm{Vac}\right\rangle ,
\end{equation}%
with eigenenergy%
\begin{equation}
E(n,0)=2\sum_{\left\{ j\right\} }\varepsilon _{j}
\end{equation}%
where $\left\{ j\right\} $\ is an arbitrary set of energy level indices. We
simply have 
\begin{eqnarray}
\left\vert \psi (n-1,-1/2)\right\rangle &=&A_{j_{0},\uparrow }\left\vert
\psi (n,0)\right\rangle  \notag \\
\left\vert \psi (n-1,1/2)\right\rangle &=&A_{j_{0},\downarrow }\left\vert
\psi (n,0)\right\rangle ,
\end{eqnarray}%
and%
\begin{eqnarray}
\left\vert \psi (n-2,0)\right\rangle &=&A_{j_{0},\uparrow }\left\vert \psi
(n-1,1/2)\right\rangle ,  \notag \\
&=&-A_{j_{0},\downarrow }\left\vert \psi (n-1,-1/2)\right\rangle ,
\end{eqnarray}%
for arbitrary $j_{0}\in \left\{ j\right\} $. Accordingly, we have 
\begin{eqnarray}
E(n-1,\pm 1/2) &=&2\sum_{\left\{ j\right\} }\varepsilon _{j}-\varepsilon
_{j_{0}},  \notag \\
E(n-2,0) &=&2\sum_{\left\{ j\right\} }\varepsilon _{j}-2\varepsilon _{j_{0}}.
\end{eqnarray}%
Then taking $\mu _{\uparrow }=\mu _{\downarrow }=\mu =\varepsilon _{j_{0}}$\
the resonant subspace can be constructed with $\kappa _{1}=\kappa _{2}$ and $%
U_{\mathrm{b}}=0$. We conclude that complete pair transport can occur for
an arbitrary pair in the zero $U$ system.

For nonzero $U$, in general, we have $\kappa _{1}\neq \kappa _{2}$, which
results in incomplete extraction. However, it is presumably that we have $%
\kappa _{1}\approx \kappa _{2}$\ for small $U$. To verify the above
analysis, numerical simulations for a finite $1$D system with different
fillings $n$ and values of $U_{l}=U$ are performed. The lattice is
illustrated in Fig. \ref{fig1}. The initial state is prepared as $\left\vert
\psi (t=0)\right\rangle =\left\vert \psi _{\mathrm{g}}(n,0)\right\rangle
\left\vert 0\right\rangle _{\mathrm{p}}$, and the evolved state is
calculated as $\left\vert \psi (t)\right\rangle =e^{-iHt}\left\vert \psi _{%
\mathrm{g}}(t=0)\right\rangle /\left\vert e^{-iHt}\left\vert \psi _{\mathrm{g%
}}(t=0)\right\rangle \right\vert $. Here the numerical computations are
performed by using a uniform mesh in conducting time discretization. In Fig. %
\ref{fig2}, we show the number of electron pairs of the side-coupled site as
a function of time, which is defined as 
\begin{equation}
\mathcal{N}_{\mathrm{p}}(t)=\left\langle \psi (t)\right\vert n_{\mathrm{p}%
,\uparrow }n_{\mathrm{p}\mathbf{,}\downarrow }\left\vert \psi
(t)\right\rangle .
\end{equation}%
We can see that, for small $J_{\mathrm{p}}$, the electron pair oscillates
between the Hubbard chain and the side-coupled site completely. When $J_{%
\mathrm{p}}$ and $U$ get large, the dynamics of electron pair oscillation
become imperfect. It follows that two ground states $\left\vert \psi _{%
\mathrm{g}}(n,0)\right\rangle $ and $\left\vert \psi _{\mathrm{g}%
}(n-2,0)\right\rangle $ are connected by a pair of electrons with binding
energy $U_{\mathrm{b}}$. It has the implication that a correlated ground
state can emit or absorb a bound pair of electrons coherently.

\begin{figure*}[tbh]
\centering
\includegraphics[width=1\textwidth]{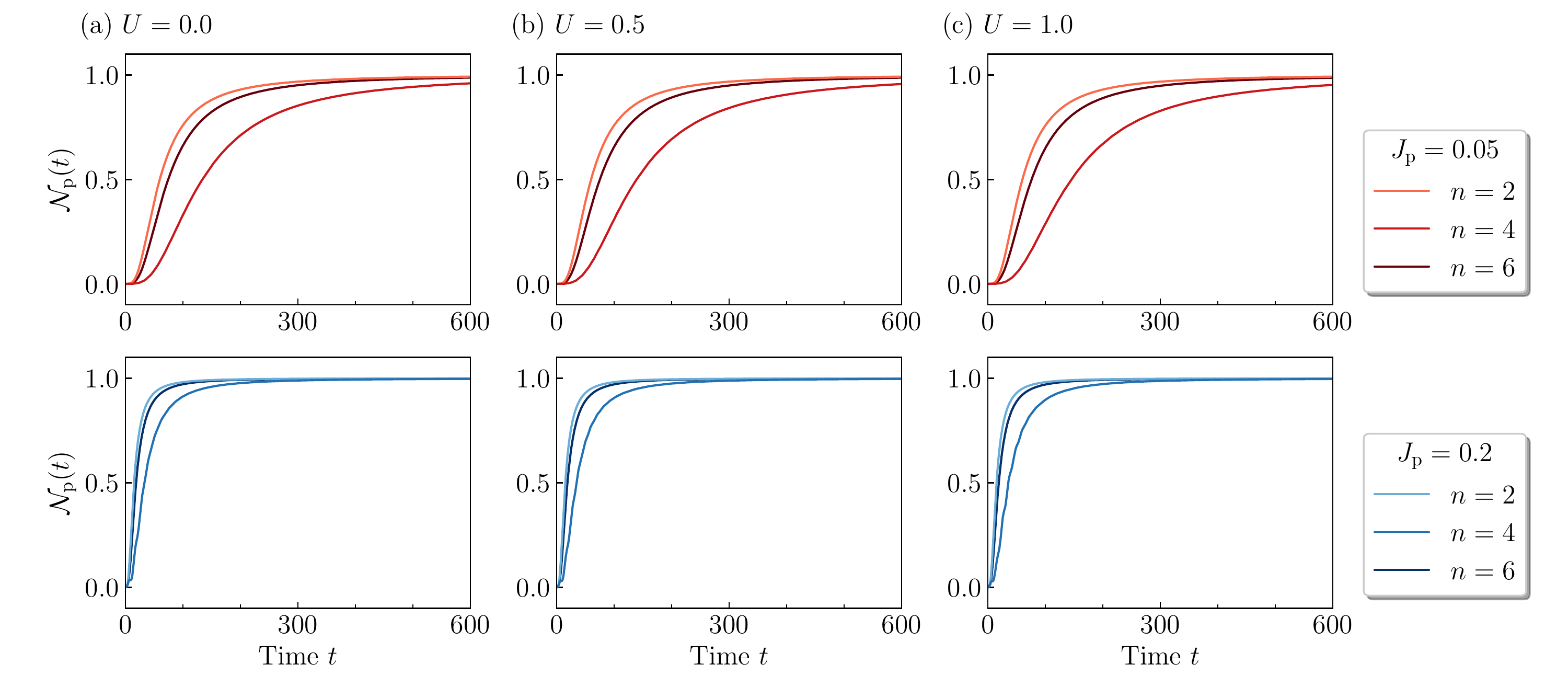}
\caption{Numerical results of the complete pair transport for different
fillings $n$ and values of $U$: (a) $U=0$, (b) $U=0.5$, and (c) $U=1.0$. The
top and bottom panels show the number of electron pairs of the side-coupled
site as a function of time $t$ for unidirectional coupling strength $J_{%
\mathrm{p}}=0.05$ and $0.2$, respectively. The parameters of the system and
the initial state are the same as that in Fig. \protect\ref{fig2}, except
that the coupling between the center system and the side-coupled site is
taken as unidirectional.}
\label{fig3}
\end{figure*}

\section{Complete pair transport}

\label{Complete pair transport}

In the above section, we have shown that a single electron can be extracted
completely from the ground state of a Hubbard model, while partially for a
pair of electrons. A natural question is whether one can realize a complete
pair extraction via another setup. We reconsider this issue by a non-Hermitian
tunneling between $H_{\mathrm{c}}$\ and the side-coupled site. The new
version of the Hamiltonian is in the form%
\begin{equation}
\mathcal{H}=H_{\mathrm{c}}+\mathcal{H}_{\mathrm{p}},
\end{equation}%
where the non-Hermitian side-coupled term is 
\begin{equation}
\mathcal{H}_{\mathrm{p}}=H_{\mathrm{p}}-J_{\mathrm{p}}\sum_{\sigma =\uparrow
,\downarrow }c_{\alpha ,\sigma }^{\dagger }c_{\mathrm{p},\sigma }
\end{equation}%
representing unidirectional tunneling. Under the resonant condition, the
corresponding matrix representation becomes%
\begin{equation}
\hslash _{\text{\textrm{R}}}=\left( 
\begin{array}{cccc}
0 & 0 & 0 & 0 \\ 
\kappa _{1} & 0 & 0 & 0 \\ 
\kappa _{1} & 0 & 0 & 0 \\ 
0 & \kappa _{2} & \kappa _{2} & 0%
\end{array}%
\right) +E_{\mathrm{g}}(n,0).
\end{equation}%
Its Jordan form contains a Jordan block of order three and the coalescing
state is $\left\vert 4\right\rangle =\left\vert \psi _{\mathrm{g}%
}(n-2,0)\right\rangle \left\vert \uparrow \downarrow \right\rangle _{\mathrm{%
p}}$. The EP dynamics allows a particular time evolution of the initial
state $\left\vert \psi _{\mathrm{g}}(n,0)\right\rangle \left\vert
0\right\rangle _{\mathrm{p}}$, that is 
\begin{eqnarray}
&&\left\vert \psi (t)\right\rangle =e^{-i\hslash _{\text{\textrm{R}}%
}t}\left\vert 1\right\rangle  \notag \\
&=&\left\vert 1\right\rangle -i\kappa _{1}t\left( \left\vert 2\right\rangle
+\left\vert 3\right\rangle \right) -t^{2}\kappa _{1}\kappa _{2}\left\vert
4\right\rangle ,
\end{eqnarray}%
which results in a steady final state $\left\vert \psi _{\mathrm{g}%
}(n-2,0)\right\rangle \left\vert \uparrow \downarrow \right\rangle _{\mathrm{%
p}}$\ at a time $t\left\vert \kappa _{1}\kappa _{2}\right\vert \gg
\left\vert \kappa _{1}\right\vert $. Importantly, unlike the Hermitian
system, the result of complete pair extraction is not sensitive to the
relation between $\kappa _{1}$ and $\kappa _{2}$.

In Fig. \ref{fig3}, we shown the numerical results of this process for
a finite system with different fillings $n$ and values of $U$. To facilitate
comparison, the parameters of the system and the initial state are the same
as that in Fig. \ref{fig2}, except that the coupling between the center
system and the side-coupled site is taken as unidirectional. We can see that, 
for a larger $J_{\mathrm{p}}$, the evolved state reaches the steady state
faster, which is an advantage in comparison to the Hermitian case in Fig. %
\ref{fig2}.

Intuitively, the result seems to be straightforward due to the
unidirectional hopping term in $\mathcal{H}_{\mathrm{p}}$. We would like to
point out that\ the resonant condition is necessary for the complete
extraction. A simple derivation can show that any deviation from the
resonant condition will result in periodic evolution rather than a steady
final state.

\section{Time-ordered bound-pair injection}

\label{Time-ordered bound-pair injection}

\begin{figure}[t]
\centering
\includegraphics[width=0.5\textwidth]{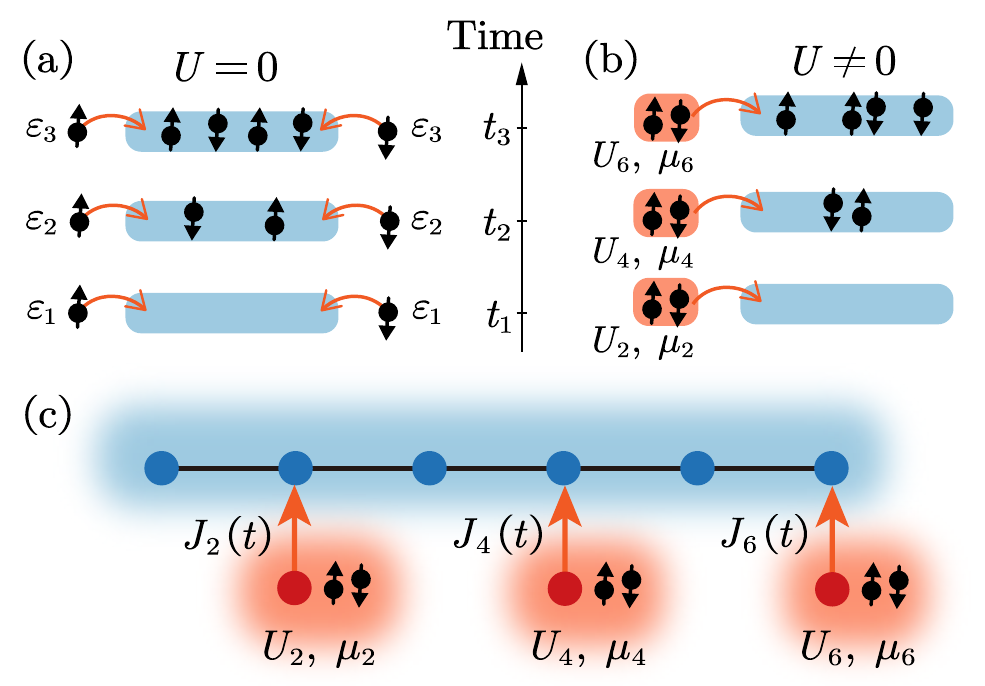}  
\caption{Schematic of the time-ordered injection process. To reach the
ground state of the system, (a) when $U=0$, the electrons can be injected
into the system one by one, without the need of specific order, (b) while
for the case of $U\neq 0$, the electrons must be filled into the system pair
by pair with order. (c) Configuration of the system and the initial state.
The system includes a six-site Hubbard chain and three side-coupled sites with
different pair binding energy and chemical potential determined by Eq. (\protect\ref{U2l}). The unidirectional coupling $J_{2}$, $J_{4} $, and $J_{6}$
(orange arrows) are switched on at ordered times $t_1$, $t_2$, and $t_3$,
respectively.}
\label{fig4}
\end{figure}

In this section, we focus on the possibility to build a ground state of the 
Hubbard model from an empty system by multipair injections. We consider a
Hubbard system with multi-side-coupled sites, which is schematically
illustrated in Fig. \ref{fig4}(c). The setup consists of two parts: $N$%
-site Hubbard model as a central system and $N/2$-site decoupled Hubbard
model as a reservoir system. The Hamiltonian reads%
\begin{equation}
H_{1}=H_{\mathrm{c}}+H_{\mathrm{r}}+H_{\mathrm{q}},
\end{equation}%
Here $H_{\mathrm{c}}$ is taken as the Hubbard chain with uniform hopping $%
J_{l,l+1}=J_{\mathrm{c}}$ and interaction $U_{l}=U$; $H_{\mathrm{r}}$ is the
reservoir term 
\begin{equation}
H_{\mathrm{r}}=\sum_{l=1}^{N/2}[\mu _{2l}(d_{2l,\uparrow }^{\dag
}d_{2l,\uparrow }+d_{2l,\downarrow }^{\dag }d_{2l,\downarrow
})+U_{2l}d_{2l,\uparrow }^{\dag }d_{2l,\uparrow }d_{2l,\downarrow }^{\dag
}d_{2l,\downarrow }],
\end{equation}%
and the quenched term is%
\begin{equation}
H_{\mathrm{q}}=\sum_{l=1}^{N/2}\sum_{\sigma =\uparrow ,\downarrow
}J_{2l}(t)c_{2l,\sigma }^{\dag }d_{2l,\sigma },
\end{equation}%
where $c_{l,\sigma }$ and $d_{l,\sigma }$\ are fermion operators and $%
J_{2l}(t)$\ is time dependent. Here both $H_{\mathrm{c}}$\ and $H_{\mathrm{r}%
}$ are Hermitian, describing the central system and reservoir system,
respectively. Notably, $H_{\mathrm{q}}$\ is a non-Hermitian term,\
representing the connection between two systems\ $H_{\mathrm{c}}$\ and $H_{%
\mathrm{r}}$. The set of parameters $\left\{ \mu _{2l},U_{2l}\right\} $ ($%
l\in \lbrack 1,N/2]$) are determined by 
\begin{eqnarray}
\mu _{2l} &=&E_{\mathrm{g}}(2l,0)-E_{\mathrm{g}}(2l-1,1/2),  \notag \\
U_{2l} &=&2E_{\mathrm{g}}(2l-1,1/2)-E_{\mathrm{g}}(2l-2,0)-E_{\mathrm{g}%
}(2l,0),  \label{U2l}
\end{eqnarray}%
where $E_{\mathrm{g}}(L,s_{z})$ is the ground-state energy of $H_{\mathrm{c}%
} $ in $L$ electron and $s_{z}$\ invariant subspace, i.e., $H_{\mathrm{c}%
}\left\vert \psi _{\mathrm{g}}(L,s_{z})\right\rangle =E_{\mathrm{g}%
}(L,s_{z})\left\vert \psi _{\mathrm{g}}(L,s_{z})\right\rangle $. In this
work, $J_{2l}(t)$\ is taken in step functions of time%
\begin{equation}
J_{2l}(t)=\left\{ 
\begin{array}{cc}
0, & t\leq\left( l-1\right) \tau \\ 
J & t>\left( l-1\right) \tau%
\end{array}%
\right. .  \label{J2l}
\end{equation}%
Obviously, $H_{\mathrm{q}}$\ is the term for a sequence of quenching with an
interval $\tau $.

Based on the above analysis, the dynamics of $H_{1}$ is governed by the time
evolution operator%
\begin{equation}
U(t)=\exp [-i\int_{0}^{t}H(t^{\prime })\text{\textrm{d}}t^{\prime }].
\label{U(t)}
\end{equation}%
Initially, the whole system is prepared in the state with $H_{\mathrm{c}}$
being empty while $H_{\mathrm{r}}$ is fully filled, that is 
\begin{equation}
\left\vert \psi (0)\right\rangle =\prod_{l=1}^{N/2}d_{2l,\uparrow }^{\dag
}d_{2l,\downarrow }^{\dag }\left\vert \mathrm{Vac}\right\rangle .
\end{equation}%
It is expected that the evolved states at $t=n\tau $ satisfy%
\begin{equation}
\left. 
\begin{array}{c}
\left\vert \psi (\tau )\right\rangle =\left\vert \psi _{\mathrm{g}%
}(2,0)\right\rangle \prod_{l=2}^{N/2}d_{2l,\uparrow }^{\dag
}d_{2l,\downarrow }^{\dag }\left\vert \mathrm{Vac}\right\rangle , \\ 
\vdots \\ 
\left\vert \psi (n\tau )\right\rangle =\left\vert \psi _{\mathrm{g}%
}(2n,0)\right\rangle \prod_{l=n+1}^{N/2}d_{2l,\uparrow }^{\dag
}d_{2l,\downarrow }^{\dag }\left\vert \mathrm{Vac}\right\rangle , \\ 
\vdots \\ 
\left\vert \psi (\infty )\right\rangle =\left\vert \psi _{\mathrm{g}%
}(N,0)\right\rangle .%
\end{array}%
\right\}
\end{equation}%
Ideally, it follows that the ground state at half filling is achieved via a
sequence of quenching.

It should be noted that the set of parameters $\left\{ \mu _{2l}\right\} $
obey the order $\mu _{2}<...<\mu _{2l}<...<\mu _{N}$,\ and $\left\{
J_{2l}(t)\right\} $ match this order. When $U=0$, $\mu _{2l}$\ reduces to $%
\varepsilon _{l}$. Remarkably, the order in $\left\{ J_{2l}(t)\right\} $\
becomes not necessary, i.e., the corresponding ground state can be built by
injecting electrons in an arbitrary way [see Figs. \ref{fig4}(a) and \ref{fig4}(b)].
This is a direct reflection of the difference between the ground states of
correlated and non-interacting systems.

\begin{figure*}[tbh]
\centering
\includegraphics[width=1\textwidth]{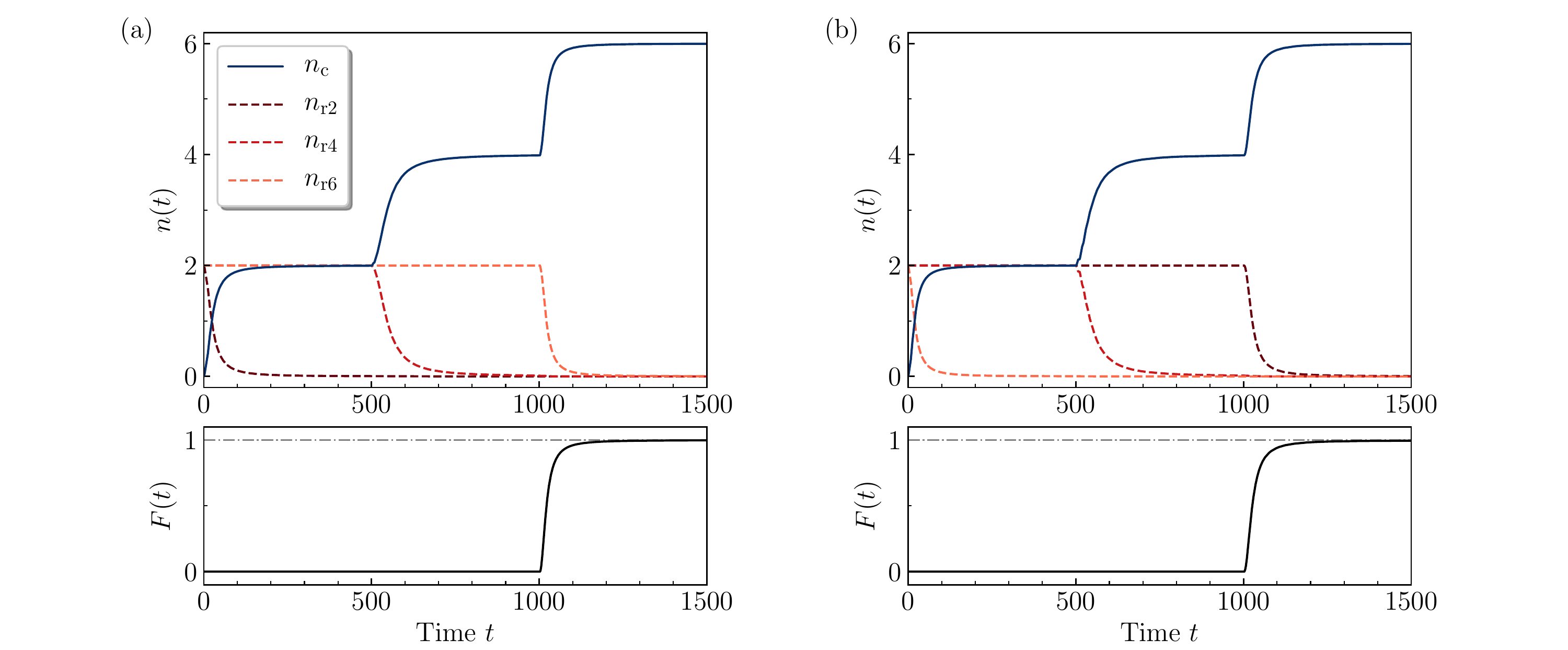}  
\caption{Numerical results of the particle density and the fidelity for the
formation processes of the ground state at half filling. Here $n_{\mathrm{c}
} $ [Eq. (\protect\ref{nc})] and $n_{\mathrm{r}2l}$ [Eq. (\protect\ref{nr2l}
)] represent the particle density for the center system and reservoir sites,
respectively. The bottom panels show the fidelity $F(t)$ [Eq. (\protect\ref%
{Fidelity})] between the evolved state and the target ground state. The
system configuration and the initial state are illustrated in Fig. \protect
\ref{fig4}(c), where the center system is empty and each side-coupled site
has two electron filled. (a) Formation processes of the ground state for
system with interaction strength $U=1$. The time dependent unidirectional
coupling $J_{2l}(t)$ is taken as the form in Eq. (\protect\ref{J2l}). (b)
When $U=0$, the order in $\left\{J_{2l}(t)\right\}$ becomes not necessary.
In this case, $J_{2l}(t)$ is taken as another order in Eq. (\protect\ref%
{J2l_v2}). Other parameters of the system: $J_{\mathrm{c}}=1$, $N=6 $, $%
J=0.1 $, and $\protect\tau=501$.}
\label{fig5}
\end{figure*}

Numerical simulations for the formation processes of the ground state at
half filling are performed for a finite system with $U=1$ and $U=0$. We
calculate the particle density for the center system and the reservoir as a
function of time, which are defined as 
\begin{equation}
n_{\mathrm{c}}=\sum_{l=1}^{N}\sum_{\sigma =\uparrow ,\downarrow
}\left\langle \psi (t)\right\vert c_{l,\sigma }^{\dagger }c_{l,\sigma
}\left\vert \psi (t)\right\rangle ,  \label{nc}
\end{equation}%
and%
\begin{equation}
n_{\mathrm{r}2l}=\sum_{\sigma =\uparrow ,\downarrow }\left\langle \psi
(t)\right\vert d_{2l,\sigma }^{\dag }d_{2l,\sigma }\left\vert \psi
(t)\right\rangle ,  \label{nr2l}
\end{equation}%
as well as the fidelity between the evolved state and the target ground
state 
\begin{equation}
F\left( t\right) =\left\vert \left\langle \psi (t)\right. \left\vert \psi _{%
\mathrm{g}}(N,0)\right\rangle \right\vert .  \label{Fidelity}
\end{equation}%
Here the target ground state used is obtained by exact diagonalization. In
Fig. \ref{fig5}(a), the interaction strength of the system is set as $U=1$
and the unidirectional coupling $J_{2l}(t)$ is taken as the form in Eq. (\ref%
{J2l}). We can see that the results are in accord with our prediction, while,
for the noninteracting case of $U=0$, we consider the parameter of
quenching with%
\begin{equation}
J_{2l}(t)=\left\{ 
\begin{array}{cc}
0, & t\leq\left( 3-l\right) \tau \\ 
J & t>\left( 3-l\right) \tau%
\end{array}%
\right. ,  \label{J2l_v2}
\end{equation}%
as an example, which sets a quench in an inverse order. In Fig. \ref{fig5}%
(b), it is demonstrated that the ground state can be well built, indicating
that the order in $\left\{ J_{2l}(t)\right\} $ becomes not necessary for
the noninteracting system. In addition, we note that each binding energy $%
U_{2l} $ is always negative in the sample we considered. It becomes positive
when we consider negative $U$. The implication of the observation cannot be
explained at this stage due to the limitation of the sample size.

\begin{figure*}[tbh]
\centering
\includegraphics[width=1\textwidth]{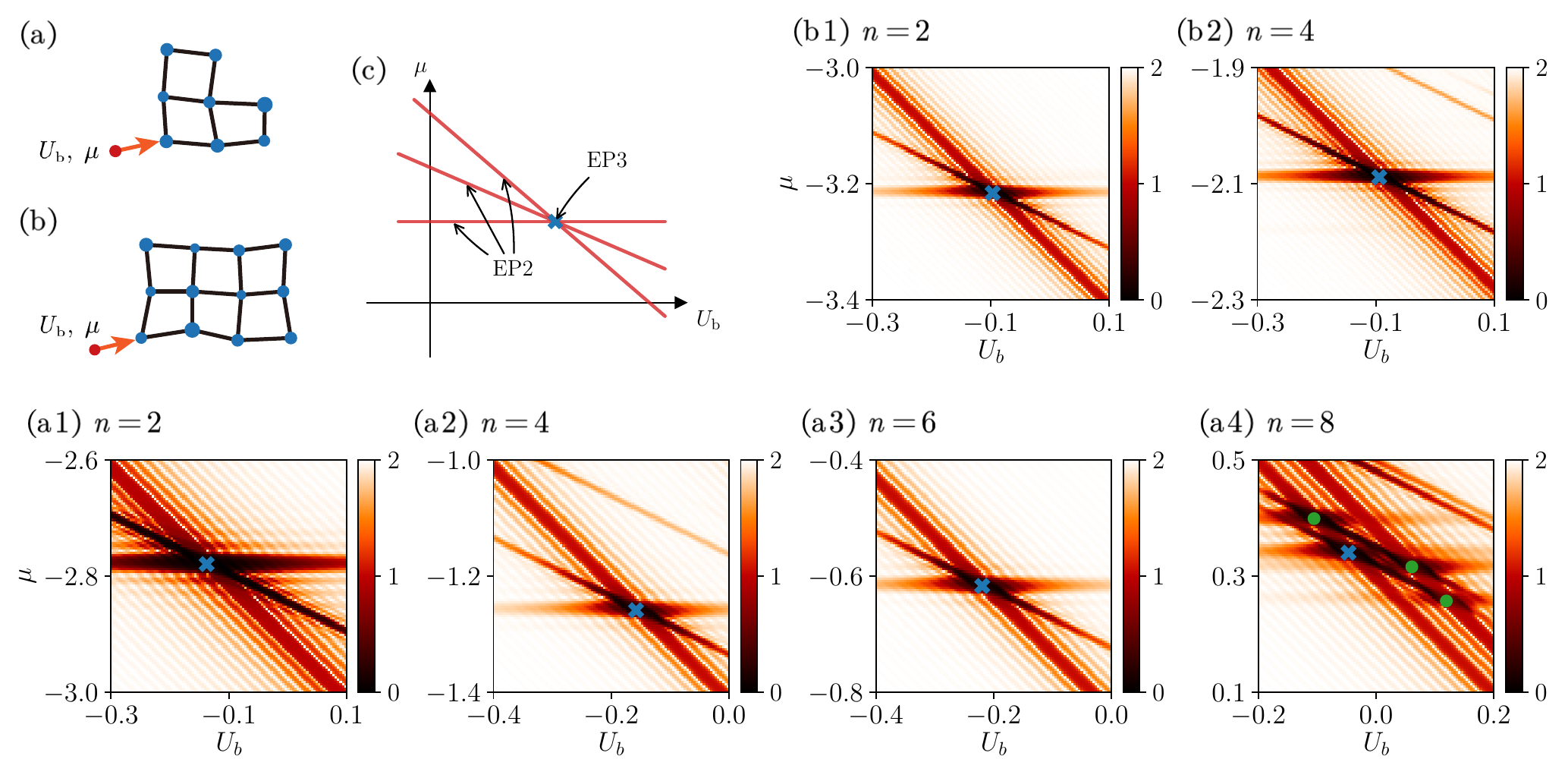}  
\caption{Panels (a) and (b) are schematics of two 2D lattices. The different
lengths of black edges and different sizes of blue solid circles denote the
disordered hopping strength $\{J_{l,l^{\prime }}\}$ and interaction strength 
$\{U_{l}\}$. (c) Schematic of the EP2 lines defined in Eqs. (\protect\ref%
{EP2_line1})-(\protect\ref{EP2_line3}), the intersection point (EP3) of
which is the resonance condition. Panels (a1)-(a4) and (b1), (b2) are numerical
results of the particle density of the side-coupled site for lattices (a) and
(b), respectively, with different number of electrons $n$. For each $n$, the
numerical simulations of time evolution are performed $100\cross100$ times
in a uniform 2D mesh of $(U_{\mathrm{b}},\protect\mu )$. The evolved time of
the final states is $t=300$ for each realization of time evolution. The blue
crosses represent values of resonance parameters obtained from exact
diagonalization. The green solid circles in (a4) are calculated from the
excited states. Other parameters of the system: $N=8$ and $12$ for (a) and
(b), and $J_{\mathrm{p}}=0.05$.}
\label{fig6}
\end{figure*}

\section{Determinations of quenching parameters and ground state}

\label{Determinations}

We have demonstrated that the ground state of the Hubbard chain can be built by
the quenching under the resonance condition (quenching parameters) in Eq. (%
\ref{U2l}). However, for a more realistic system, these parameters as well
as the ground-state energy are actually unknown. A problem arises of 
how to implement the quenching protocol without the prior knowledge of the
Hamiltonian and the quenching parameters. In this section, we show that this
problem can be solved by scanning over the parameters of the auxiliary site
and observing the number of electrons in it.

Consider the following Hamiltonian: 
\begin{equation}
H_{2}=H_{\mathrm{c}}+H_{\mathrm{p}}^{^{\prime }},  \label{H2}
\end{equation}%
with $H_{\mathrm{c}}$ defined in Eq. (\ref{Hc}), and the non-Hermitian
side-coupled term is 
\begin{equation}
H_{\mathrm{p}}^{\prime }=J_{\mathrm{p}}\sum_{\sigma =\uparrow ,\downarrow
}c_{\alpha ,\sigma }^{\dagger }c_{\mathrm{p},\sigma }+U_{\mathrm{b}}n_{%
\mathrm{p},\uparrow }n_{\mathrm{p}\mathbf{,}\downarrow }+\mu \sum_{\sigma
=\uparrow ,\downarrow }n_{\mathrm{p},\sigma }.
\end{equation}%
Here $\alpha $ denotes a site of $H_{\mathrm{c}}$ and the parameters $%
\{J_{l,l^{\prime }},U_{l}\}$ in $H_{\mathrm{c}}$ are assumed to be unknown:
in the following numerical calculation, $\{J_{l,l^{\prime }}\}$ and $%
\{U_{l}\}$ are taken as the random samples that are uniformly distributed
over the intervals $[1,1.2) $ and $[0.9,1.1)$, respectively. The system
configurations considered are schematically illustrated in Figs. \ref{fig6}%
(a) and \ref{fig6}(b).

Similarly with the analysis for the Hermitian system in Sec. \ref{Model and
single-pair oscillation}, under the basis defined in Eq. (\ref{basis}), the
effective Hamiltonian for $H_{2}$ can be written as 
\begin{equation}
\hslash =\left( 
\begin{array}{cccc}
\epsilon _{1} & \kappa _{1} & \kappa _{1} & 0 \\ 
0 & \epsilon _{2} & 0 & \kappa _{2} \\ 
0 & 0 & \epsilon _{2} & \kappa _{2} \\ 
0 & 0 & 0 & \epsilon _{4}%
\end{array}%
\right) ,
\end{equation}%
where the matrix elements 
\begin{eqnarray}
\epsilon _{1} &=&E_{\mathrm{g}}(n,0),  \notag \\
\epsilon _{2} &=&E_{\mathrm{g}}(n-1,1/2)+\mu ,  \notag \\
\epsilon _{4} &=&E_{\mathrm{g}}(n-2,0)+2\mu +U_{\mathrm{b}},  \notag \\
\kappa _{1} &=&\langle 1|H_{\mathrm{p}}^{\prime }|2\rangle =\langle 1|H_{%
\mathrm{p}}^{\prime }|3\rangle ,  \notag \\
\kappa _{2} &=&\langle 2|H_{\mathrm{p}}^{\prime }|4\rangle =\langle 3|H_{%
\mathrm{p}}^{\prime }|4\rangle .
\end{eqnarray}%
We note that, when varying the parameters $(U_{\mathrm{b}},\mu )$ of the
side-coupled site, the matrix $\hslash $ supports three sets of two-state
coalescence points (EP2), forming three lines in the parameter space of $%
(U_{\mathrm{b}},\mu )$. The equations of three lines respectively have the
forms 
\begin{eqnarray}
&&\mu =E_{\mathrm{g}}(n,0)-E_{\mathrm{g}}(n-1,1/2),  \label{EP2_line1} \\
&&U_{\mathrm{b}}+\mu = E_{\mathrm{g}}(n-1,1/2)-E_{\mathrm{g}}(n-2,0),
\label{EP2_line2} \\
&&U_{\mathrm{b}}+2\mu =E_{\mathrm{g}}(n,0)-E_{\mathrm{g}}(n-2,0),
\label{EP2_line3}
\end{eqnarray}
wherein the matrix $\hslash $ can be brought into Jordan form that contains
a Jordan block of order two by the Jordan decomposition. The intersection
point of the three lines is a three-state coalescence point (EP3), which is
also the resonance condition in Eq. (\ref{U2l}). The schematic illustration
of the three lines of EP2 and point of EP3 is given in Fig. \ref{fig6}(c).
We would like to point out that this feature is independent of the dimension
and the exact knowledge of parameters $\{J_{l,l^{\prime }},U_{l}\}$ of the
Hubbard model.

The above analysis suggests a scanning scheme to determine the quenching
parameters and the ground-state energy of the system. As we have shown in
the previous section, the system with parameters at EP supports
unidirectional dynamics. In the current system, it can be checked that, if
the initial state is prepared as $\left\vert 4\right\rangle =\left\vert \psi
_{\mathrm{g}}(n-2,0)\right\rangle \left\vert \uparrow \downarrow
\right\rangle _{\mathrm{p}}$, then after the quench and for a large $t$, the
pair probability at the side-coupled site is dominated by the factors $t^{-2}
$ and $t^{-4}$ for parameters $(U_{\mathrm{b}},\mu )$ of EP2 and EP3,
respectively. That is, the numbers of electrons at the side-coupled site
decay at different rates for parameters of EP2 and EP3. For the case with $%
(U_{\mathrm{b}},\mu )$ deviating from EPs, the electrons oscillate between
the side-coupled site $\mathrm{p}$ and the center system $H_{\mathrm{c}}$.
Scanning over $(U_{\mathrm{b}},\mu )$ for a set of realizations of time
evolution, the pattern in Fig. \ref{fig6}(c) is expected to be obtained if
we observe the evolved-state particle density for the side-coupled site. The
initial state $\left\vert 4\right\rangle$ with number of electrons $n$ can
be built by using the resonance conditions for $n-2$, $n-4$, $n-6$, etc.,
which are detected from previous scans. Then, using the scanning scheme, the
resonance conditions for $n=2,4,6,...$ can be obtained successively.

\begin{table}[t]
\caption{$(U_\mathrm{b}, \protect\mu, E_{\mathrm{g}})$ obtained from the
scanning process and $(U_\mathrm{b}^{\prime}, \protect\mu^{\prime}, E_{ 
\mathrm{g}}^{\prime})$ obtained from exact diagonalization for lattices (a)
and (b) at different number of filled electrons $n$.}
\label{table}
\centering
\begin{ruledtabular}
		\renewcommand{\arraystretch}{1.4}
		\begin{tabular}{crrrrrr}
			\multicolumn{1}{c}{(a)}  & \multicolumn{3}{c}{Scan} & \multicolumn{3}{c}{Diagonalization}\\
			\cmidrule{1-1}\cmidrule{2-4}\cmidrule{5-7}
			$n$ & $U_\mathrm{b}$ & $\mu$ & $E_{\mathrm{g}}$ & $U_\mathrm{b}^{\prime}$ & $\mu^{\prime}$ & $E_{\mathrm{g}}^{\prime}$\\
			\hline
			$2$	
			& $-0.167$ & $-2.770$ & $-5.706$ & $-0.138$ & $-2.779$ & $-5.696$\\
			\hline
			$4$ 
			& $-0.133$ & $-1.267$ & $-8.373$ &  $-0.157$ & $-1.259$ & $-8.371$\\
			\hline
			$6$
			& $-0.194$ & $-0.626$ & $-9.819$ & $-0.220$ & $-0.617$ & $-9.824$\\
			\hline
			$8$ 	
			& $-0.018$ & $0.330$ & $-9.177$ & $-0.047$ & $0.341$ & $-9.190$\\
		\end{tabular}\\[2ex]
		\begin{tabular}{crrrrrr}
			\multicolumn{1}{c}{(b)}  & \multicolumn{3}{c}{Scan} & \multicolumn{3}{c}{Diagonalization}\\
			\cmidrule{1-1}\cmidrule{2-4}\cmidrule{5-7}
			$n$ & $U_\mathrm{b}$ & $\mu$ & $E_{\mathrm{g}}$ & $U_\mathrm{b}^{\prime}$ & $\mu^{\prime}$ & $E_{\mathrm{g}}^{\prime}$\\
			\hline
			$2$	
			& $-0.078$ & $-3.222$ & $-6.522$ & $-0.097$ & $-3.216$ & $-6.528$ \\
			\hline
			$4$ 
			& $-0.070$ & $-2.100$ & $-10.788$ & $-0.094$ & $-2.089$ & $-10.800$ \\
		\end{tabular}
	\end{ruledtabular}
\end{table}

\begin{figure}[tbh]
\centering
\includegraphics[width=0.45\textwidth]{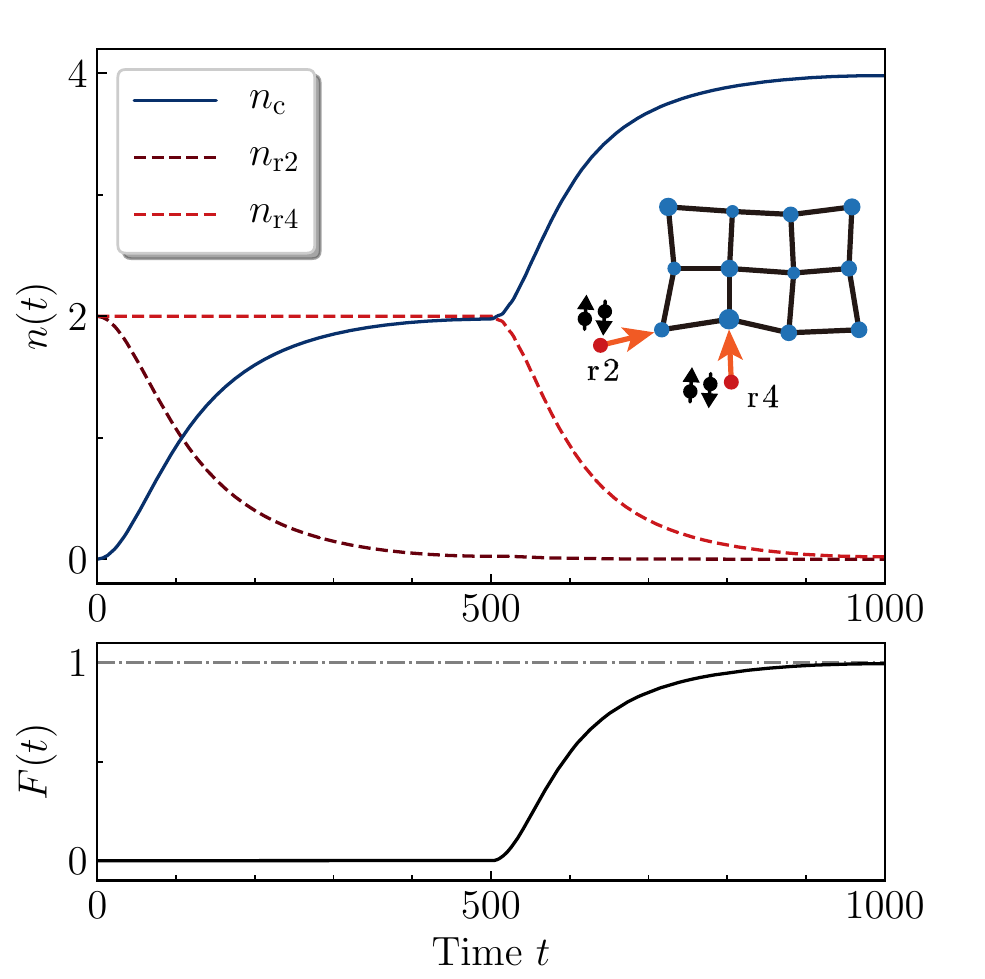}  
\caption{Numerical results of the particle density and the fidelity for the
formation processes of the ground state at filling $n=4$. Here $n_{\mathrm{c}
} $ and $n_{\mathrm{r}2l}$ represent the particle density for the center
system and reservoir sites, respectively. The inset illustrates the system
configuration and the initial state. The bottom panel shows the fidelity $%
F(t)$ between the evolved state and the target ground state [four-electron
ground state $\left\vert \protect\psi _{\mathrm{g}}(n=4,s^{z}=0)\right\rangle
$ obtained from exact diagonalization]. The disordered parameters $%
\{J_{l,l^{\prime }},U_{l}\}$ are taken to be the same as that of Figs. 
\protect\ref{fig6}(b1) and \protect\ref{fig6}(b2). The quenching parameters $(U_\mathrm{b}, 
\protect\mu)$ with $n=2$ and $4$ are taken as the scanning values in Table 
\protect\ref{table}(b). Other parameters of the system: $N=12 $, $J_{\mathrm{%
p}}=0.05 $, and $\protect\tau=501$.}
\label{fig7}
\end{figure}

To verify the performance of the scanning scheme, we conduct numerical
simulations with initial state $\left\vert \psi(0)\right\rangle =\left\vert
\psi _{\mathrm{g}}(n-2,0)\right\rangle \left\vert \uparrow \downarrow
\right\rangle _{\mathrm{p}}$ for different numbers of electrons $n$, and
evolve the state under the Hamiltonian in Eq. (\ref{H2}), with
aforementioned disordered parameters $\{J_{l,l^{\prime }},U_{l}\}$. For each 
$n$, we scan over the parameters of the side-coupled site $(U_{\mathrm{b}%
},\mu )$ by $100\cross100$ realizations of time evolution. In practical
implementation, one only needs to scan two lines in the parameter space (for
instance, two lines with constant $\mu$) to obtain a few points of EP2, and
then, connecting them in straight lines, the intersection is the position of the 
resonance point (EP3). For each realization, the system is reprepared in
state $\left\vert \psi(0)\right\rangle$ initially. After a sufficiently long
evolved time, we calculate particle density for the side-coupled site. The
calculations of time evolution are performed in the truncated subspace of
around $100$ low-excited states, since the contribution of the high-excited
states are very tiny to the quenching process.

The numerical results are presented in Figs. \ref{fig6}(a1)-\ref{fig6}(a4) and 
\ref{fig6}(b1), \ref{fig6}(b2) for the lattices in Figs. \ref{fig6}(a) and \ref{fig6}(b), respectively. We
can clearly see the patterns of EP lines for each $n$. The blue crosses
represent values of resonance parameters obtained from exact diagonalization
and Eq. (\ref{U2l}), which are well located at the intersection of the EP
lines obtained from the scanning numerical simulations. In Fig. \ref{fig6}%
(a4), more intersection points that come from the low-excited states are
obtained. This is verified by the green solid circles that are calculated
from Eq. (\ref{U2l}) but changing $E_{\mathrm{g}}(n,0)$ or $E_{\mathrm{g}%
}(n-1,1/2)$ to the corresponding first-excited energies. In Table \ref{table}%
, we present the scanning values of $(U_\mathrm{b}, \mu)$ that are obtained from
the minimum of the particle density of a side-coupled site in the data of Figs. %
\ref{fig6}(a1)-\ref{fig6}(a4) and \ref{fig6}(b1), \ref{fig6}(b2). Using these scanning values as inputs,
the ground-state energy $E_{\mathrm{g}}$ for different $n$ is calculated
from Eq. (\ref{U2l}). By contrast, $E_{\mathrm{g}}^{\prime}$ is obtained
from exact diagonalization and $(U_\mathrm{b}^{\prime}, \mu^{\prime})$ is
calculated from Eq. (\ref{U2l}). We can see that, for both lattices, the
values of $E_{\mathrm{g}}$ and $E_{\mathrm{g}}^{\prime}$ are close at each
number of filled electron $n$.

Furthermore, we use the scanning values of quenching parameters $(U_\mathrm{b%
}, \mu)$ in Table \ref{table}(b) to build the four-electron ground states of
the $2$D lattice in Fig. \ref{fig6}(b). The quenching process is the same as
that described in Sec. \ref{Time-ordered bound-pair injection}, and the time
evolution here is calculated by the exact diagonalization of the full
Hamiltonian in four-particle subspace. With high fidelity between the target
ground state $\left\vert \psi _{\mathrm{g}}(n=4,s^{z}=0)\right\rangle$ and
the evolved state at $t=1000$, the numerical results presented in Fig. \ref%
{fig7} show the validity of the data obtained from the scanning
process, as well as verify that the quenching protocol can be applied to the $2
$D Hubbard model deviating from half filling.

\section{Summary}

\label{Summary}

In summary, we have proposed a method to build the ground state of a Hubbard
model with $n$ electrons to the one with $n+2$ electrons by injecting a
bound pair of electrons in the framework of quantum dynamics. The underlying
mechanism is the EP dynamics at the resonance. As application, we
demonstrated the dynamic formation of the ground state of the Hubbard model at
half filling and deviating from half filling. It provides a
clear picture for the correlated ground state: it can be obtained by filling
a group of bound pair electrons in time order. Each pair with
specific filling order has its own binding energy and chemical potential,
which can be determined by the proposed scanning scheme. In contrast, when
the Hubbard repulsion $U$ is zero, the binding energy vanishes, and then the
filling order is not necessary. It is expected that our finding not only 
provides a method for correlated quantum state engineering, but also reveals
the feature of the ground state in an alternative way.

\acknowledgments This work was supported by National Natural Science
Foundation of China (under Grant No. 11874225).

\section*{Appendix}

\setcounter{equation}{0} \renewcommand{\theequation}{A\arabic{equation}}

In this appendix, we present the derivations on the time evolutions under
Hermitian matrix $h_{\mathrm{R}}$\ and\ non-Hermitian matrix $\hslash _{%
\mathrm{R}}$ to explain the connection and difference between two dynamic
processes. For simplicity, we omit the diagonal elements $E_{\mathrm{g}}(n,0)
$\ in both cases.

(i) The eigenvectors and eigenvalues of matrix 
\begin{equation}
h_{\mathrm{R}}=\left( 
\begin{array}{cccc}
0 & \kappa _{1} & \kappa _{1} & 0 \\ 
\kappa _{1} & 0 & 0 & \kappa _{2} \\ 
\kappa _{1} & 0 & 0 & \kappa _{2} \\ 
0 & \kappa _{2} & \kappa _{2} & 0%
\end{array}%
\right) 
\end{equation}%
are in the form%
\begin{eqnarray}
\left\vert \phi _{0}\right\rangle  &=&\frac{\sqrt{2}}{\varepsilon }\left( 
\begin{array}{c}
-\kappa _{2} \\ 
0 \\ 
0 \\ 
\kappa _{1}%
\end{array}%
\right) ,E_{0}=0;  \notag \\
\left\vert \phi _{\pm }\right\rangle  &=&\frac{1}{\varepsilon }\left( 
\begin{array}{c}
2\kappa _{1} \\ 
\pm \varepsilon  \\ 
\pm \varepsilon  \\ 
2\kappa _{2}%
\end{array}%
\right) ,E_{\pm }=\pm \varepsilon ;  \notag \\
\left\vert \phi _{1}\right\rangle  &=&\left( 
\begin{array}{c}
0 \\ 
-1 \\ 
1 \\ 
0%
\end{array}%
\right) ,E_{1}=0;
\end{eqnarray}%
where $\varepsilon =\sqrt{2(\kappa _{1}^{2}+\kappa _{2}^{2})}$.
Straightforward derivation shows that%
\begin{equation}
\left\vert 1\right\rangle =\frac{\kappa _{1}\varepsilon \left( \left\vert
\phi _{+}\right\rangle +\left\vert \phi _{-}\right\rangle \right) -2\sqrt{2}%
\kappa _{2}\varepsilon \left\vert \phi _{0}\right\rangle }{4\left( \kappa
_{1}^{2}+\kappa _{2}^{2}\right) }
\end{equation}%
and%
\begin{eqnarray}
&&e^{-ih_{\mathrm{R}}t}\left\vert 1\right\rangle   \notag \\
&=&\frac{1}{2\left( \kappa _{1}^{2}+\kappa _{2}^{2}\right) }\left( 
\begin{array}{c}
2\kappa _{1}^{2}\cos \left( \varepsilon t\right) +2\kappa _{2}^{2} \\ 
-i\varepsilon \kappa _{1}\sin \left( \varepsilon t\right)  \\ 
-i\varepsilon \kappa _{1}\sin \left( \varepsilon t\right)  \\ 
2\kappa _{2}\kappa _{1}\cos \left( \varepsilon t\right) -2\kappa _{2}\kappa
_{1}%
\end{array}%
\right) .
\end{eqnarray}%
Then the pair probability at the side-coupled site is 
\begin{equation}
\left\vert \left\langle 4\right\vert e^{-ih_{\mathrm{R}}t}\left\vert
1\right\rangle \right\vert ^{2}=\frac{4\left( \kappa _{2}\kappa _{1}\right)
^{2}\cos ^{2}\left( \varepsilon t/2\right) }{\left( \kappa _{1}^{2}+\kappa
_{2}^{2}\right) ^{2}},
\end{equation}%
with the maximum%
\begin{equation}
\text{Max}\left\vert \left\langle 4\right\vert e^{-ih_{\mathrm{R}%
}t}\left\vert 1\right\rangle \right\vert ^{2}=\frac{4\left( \kappa
_{2}\kappa _{1}\right) ^{2}}{\left( \kappa _{1}^{2}+\kappa _{2}^{2}\right)
^{2}}.
\end{equation}%
It approaches to unit when $\kappa _{1}=\kappa _{2}$.

(ii) For the non-Hermitian matrix%
\begin{equation}
\hslash _{\mathrm{R}}=\left( 
\begin{array}{cccc}
0 & 0 & 0 & 0 \\ 
\kappa _{1} & 0 & 0 & 0 \\ 
\kappa _{1} & 0 & 0 & 0 \\ 
0 & \kappa _{2} & \kappa _{2} & 0%
\end{array}%
\right) ,
\end{equation}%
we simply have%
\begin{equation}
\left( \hslash _{\mathrm{R}}\right) ^{2}=\left( 
\begin{array}{cccc}
0 & 0 & 0 & 0 \\ 
0 & 0 & 0 & 0 \\ 
0 & 0 & 0 & 0 \\ 
2\kappa _{1}\kappa _{2} & 0 & 0 & 0%
\end{array}%
\right) 
\end{equation}%
and%
\begin{equation}
\left( \hslash _{\mathrm{R}}\right) ^{3}=\mathbf{0}.
\end{equation}%
Then we have 
\begin{equation}
e^{-i\hslash _{\mathrm{R}}t}=\left( 
\begin{array}{cccc}
1 & 0 & 0 & 0 \\ 
-it\kappa _{1} & 1 & 0 & 0 \\ 
-it\kappa _{1} & 0 & 1 & 0 \\ 
-t^{2}\kappa _{1}\kappa _{2} & -it\kappa _{2} & -it\kappa _{2} & 1%
\end{array}%
\right) ,
\end{equation}%
which results in%
\begin{equation}
e^{-i\hslash _{\mathrm{R}}t}\left\vert 1\right\rangle =\left( 
\begin{array}{c}
1 \\ 
-it\kappa _{1} \\ 
-it\kappa _{1} \\ 
-t^{2}\kappa _{1}\kappa _{2}%
\end{array}%
\right) .
\end{equation}%
It turns out that 
\begin{equation}
e^{-i\hslash _{\mathrm{R}}t}\left\vert 1\right\rangle \rightarrow \left\vert
4\right\rangle ,
\end{equation}%
for a sufficiently long time without the need of $\kappa _{1}=\kappa _{2}$.

\end{document}